\documentclass[conference]{IEEEtran}
\IEEEoverridecommandlockouts
\usepackage{cite}
\usepackage{amsmath,amssymb,amsfonts}
\usepackage{algorithmic}
\usepackage{graphicx}
\usepackage{textcomp}
\usepackage{xcolor}
\usepackage{comment}
\usepackage{booktabs} 
\usepackage{multirow}
\usepackage{subcaption} %new entry
\usepackage{braket} % for \ket and quantum notation
\usepackage{authblk} %new entry

\def\BibTeX{{\rm B\kern-.05em{\sc i\kern-.025em b}\kern-.08em
    T\kern-.1667em\lower.7ex\hbox{E}\kern-.125emX}}
\usepackage{ragged2e} % For \RaggedRight command
\usepackage{array}     % For defining new column types
\newcolumntype{L}[1]{>{\RaggedRight\hspace{0pt}}p{#1}} % New Left-aligned column type
\begin{document}

\title{Reliable Audio Deepfake Detection in Variable Conditions via Quantum‑Kernel SVMs
}

\author{
\makebox[\textwidth][c]{%
\parbox{0.9\textwidth}{\centering
Lisan Al Amin, Vandana P. Janeja\\
University of Maryland, Baltimore County, Baltimore, Maryland, USA\\
salamin1@umbc.edu; vjaneja@umbc.edu
}}
}

\maketitle

\begin{abstract}

Detecting synthetic speech is challenging when labeled data are scarce and recording conditions vary. Existing end-to-end deep models often overfit or fail to generalize, and while kernel methods can remain competitive, their performance heavily depends on the chosen kernel. Here, we show that using a quantum kernel in audio deepfake detection reduces false-positive rates without increasing model size. Quantum feature maps embed data into high-dimensional Hilbert spaces, enabling the use of expressive similarity measures and compact classifiers.
Building on this motivation, we compare quantum-kernel SVMs (QSVMs) with classical SVMs using identical mel-spectrogram preprocessing and stratified 5-fold cross-validation across four corpora (ASVspoof~2019 LA, ASVspoof~5 (2024), ADD23, and an In-the-Wild set). QSVMs achieve consistently lower equal-error rates (EER): 0.183 vs.\ 0.299 on ASVspoof~5 (2024), 0.081 vs.\ 0.188 on ADD23, 0.346 vs.\ 0.399 on ASVspoof~2019, and 0.355 vs.\ 0.413 In-the-Wild. At the EER operating point (where FPR equals FNR), these correspond to absolute false-positive-rate reductions of 0.116 (38.8\%), 0.107 (56.9\%), 0.053 (13.3\%), and 0.058 (14.0\%), respectively. We also report how consistent the results are across cross‑validation folds and margin‑based measures of class separation, using identical settings for both models. The only modification is the kernel; the features and SVM remain unchanged, no additional trainable parameters are introduced, and the quantum kernel is computed on a conventional computer.

%

%These results indicate that quantum kernels are a viable, drop-in alternative to classical kernels for small-data and variable-condition detection, improving separability and reducing false positives without increasing model size. Our study is conducted in a classical simulation, isolating the kernel’s effect, which can be integrated into established pipelines as a kernel swap.
\end{abstract}
\section{Introduction}
\begin{figure*}[htbp]
    \centering
    \includegraphics[width=0.8\textwidth]{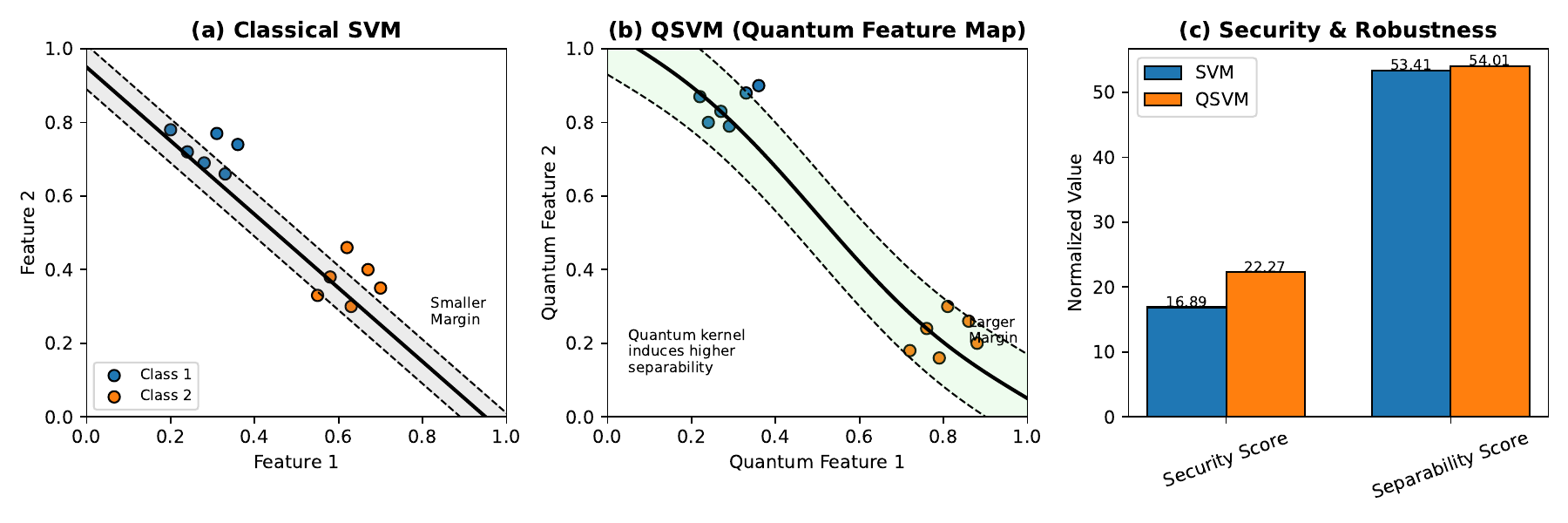}
    \caption{Comparison of feature separability and security scores across models, where panels (a) and (b) are illustrated and panel (c) is computed on ASVspoof~2019 folds. Panel (a) illustrates the decision boundary and feature separability achieved by the classical SVM, while Panel (b) depicts the enhanced margin width and separability obtained by the QSVM. Panel (c) presents a bar chart of the corresponding security and separability scores, highlighting the relative improvements. Overall, the QSVM exhibits superior performance, achieving higher separability and security compared to the classical SVM.}
    \label{fig:methodology}
\end{figure*}

The rapid advancement of artificial intelligence has significantly improved speech synthesis technologies. However, this progress has also heightened security risks, particularly with the development of audio deepfakes. Audio deepfakes are created by artificially manipulating voice signals to produce speech that is nearly indistinguishable from genuine recordings. This technology poses substantial threats to areas such as live streaming, teleconferencing, public broadcasting, and digital forensics. In real-time attack scenarios, adversaries may apply precise modifications to audio signals, such as altering waveform dynamics, adjusting pitch and formant frequencies, or embedding subtle noise patterns. These changes are designed to deceive both human listeners and automated verification systems by closely imitating natural vocal features.
Within live audio pipelines, these perturbations often manifest at the earliest stages, either at the raw input layer or during initial pre-processing, before classification or authentication occurs. These attacks exploit the vulnerabilities of modern deep models. In these systems, even the slightest changes to the input result in small, predictable changes in the output. Attackers use that predictability to make minimal edits to the audio. The edits are often too minor for people to notice, yet they can still deceive the detector, making a fake sound seem real or a real clip sound fake. These models are particularly susceptible to gradient-based adversarial attacks, and performance can degrade sharply under carefully crafted perturbations. Consequently, the challenge extends beyond simply preventing human deception: it also involves fortifying the detection system itself to maintain 
reliability under adversarial conditions.
To address these challenges, our research turns to quantum computing, which offers the potential for intrinsically superior feature separability by leveraging exponentially large Hilbert spaces, capacities that surpass those of traditional Support Vector Machines (SVMs) and classical neural architectures \cite{lin2024quantum, venegas2024quantum}. In this work we focus on reducing false positives (false acceptances), the failure mode that carries the highest cost in forensic and security settings. Our question is whether the geometry induced by a quantum kernel reduces the false-positive rate under the same preprocessing and training budget compared to a strong classical SVM. Quantum techniques such as Quantum Support Vector Machines (QSVM) and quantum kernel methods can construct large-margin hyperplanes in these high-dimensional spaces, enabling more precise discrimination between genuine and spoofed audio features (Fig \ref{fig:methodology}). This improved separability directly contributes to reduced false-positive rates, which is an essential property for high-stakes domains such as forensic analysis and cybersecurity \cite{venegas2024quantum, kawa2022defense}.

In deepfake settings, the adversary controls a differentiable synthesizer (TTS/VC) and can tune the audio to evade a differentiable detector while keeping it natural to listeners. Even without direct access to the detector, they can attack a surrogate model, and the resulting perturbations often transfer to the target system. We do not evaluate attacks here and do not claim intrinsic adversarial robustness from quantum techniques; this threat model simply motivates our focus on lowering false positives.

In this work, we conduct a systematic comparison between QSVM-based methods and classical SVM baselines, aiming to quantify improvements in feature separability, adversarial robustness, and real-time scalability. We operate in a small‑data, variable‑condition regime where margin‑based kernel methods are competitive, so a classical SVM on mel‑spectrogram features serves as a strong, data‑efficient baseline. A QSVM is the minimal extension of that baseline: it keeps the same margin‑based classifier and training recipe but swaps in a quantum kernel, changing only the similarity geometry. We compute the quantum kernel in classical simulation and pass it to the same SVM solver used by the classical runs; features, folds, and preprocessing (scaling and fold‑specific PCA) are identical (Fig.~\ref{fig:methodology2}). This design avoids confounds from model size or optimization tricks, allowing us to attribute any differences in performance to the kernel itself. Our central question is therefore simple and testable: Does a more powerful kernel lower the false‑positive rate (and EER) under matched budgets? Because the change is a drop‑in kernel swap, the approach integrates naturally into existing pipelines without increasing model size. Given this motivation and the modular nature of kernel substitution, we now turn to the specific research questions that guide our investigation.

\subsection{Research Questions}

This study is guided by the following research questions:

\begin{enumerate}
   \item How can a quantum kernel be integrated as a drop in replacement within a classical SVM pipeline to improve audio deepfake detection, with a focus on lowering the false positive rate (and EER) under matched preprocessing and training budgets?
   \item To what extent does a quantum kernel, compared to standard classical kernels, enhance feature separability and produce more stable error rates across folds, as reflected by lower EER and narrower confidence intervals?
    %\item Can quantum-secure watermarking methods be integrated into the detection pipeline to ensure forensic compliance, traceability, and legal admissibility in operational environments?
    \item How do any reductions in false positives relate to margin based separability diagnostics and ROC/DET behavior in the induced feature space?

\end{enumerate}

%\paragraph*{Paper organization.}

The remainder of this paper is organized to systematically address these questions. Section~II reviews related work and situates our study, including a brief taxonomy (Table~\ref{tab:compact_taxonomy}). Section III details the methodology and kernel swap setup; Section IV reports the main results, and Section V provides dataset-wise analysis and diagnostics. Section~VI discusses implications and limitations, Section~VII outlines future work, and Section~VIII concludes.

\section{Literature Review}

Recent research on audio deepfake detection has primarily used traditional deep learning methods, with several limitations. A recent survey~\cite{yi2023audio} categorizes audio deepfake types and methods and notes a gap in interpretability, where deep models often act as ``black boxes'' with limited explanatory power, undermining trust in security‑sensitive settings. Further work~\cite{khanjani2023} reports that detection accuracy frequently drops on unseen datasets and real‑world conditions, indicating weak generalization.

Before the experiments, a brief taxonomy (Table~\ref{tab:compact_taxonomy}) summarizes threat sources, access scenarios, representations, detector families, and evaluation metrics. This framing presents the study as a controlled kernel‑to‑kernel comparison (SVM vs.\ QSVM) under identical feature preparation and training budgets. The evaluation reports false‑positive rate (FPR) and equal‑error rate (EER) across four corpora using identical folds.

\subsection*{A. Taxonomy of Audio Deepfake Threats and Detectors}

\begin{table*}[t]
\small
\centering
\renewcommand{\arraystretch}{1.1}
\begin{tabular}{L{3.0cm} L{4.2cm} L{4.8cm} L{4.2cm}} % Use L instead of p
\hline
\textbf{Dimension} & \textbf{Sub-dimensions / Examples} & \textbf{Key questions \& metrics} & \textbf{Representative methods} \\
\hline

Problem scope & Attacks: TTS, VC, neural codecs, partial edits, replay; Contexts: LA, PA, In-the-Wild & Which attacks transfer best? How to catch partial edits? Use EER, minDCF, FPR@TPR, localization F1. & ASVspoof 2021~\cite{yamagishi2021asvspoof}, ADD 2023~\cite{yi2023add}. \\

Data / Benchmarks & Cross-corpus, multi-codec, few-/zero-shot, multilingual; Augmentation with codecs, noise, RIRs & How robust to unseen codecs? Report cross-corpus metrics, codec breakdowns. & ASVspoof 2021~\cite{yamagishi2021asvspoof}, ADD 2023~\cite{yi2023add}. \\

Model families & CNNs (RawNet2), AASIST (GAT), AST/transformers, SSL+GAT, OOD/metric, QSVM kernels & Which model balances accuracy and compute? Include size vs EER. & AASIST~\cite{jung2022aasist}, AST~\cite{gong2021ast}. \\

Robustness & Codecs, noise, channel, reverb, adversarial perturbations & How does each stressor affect error? Report per-condition metrics. & Augmentation pipelines~\cite{tak2022asvspoof}, adversarial training. \\

Quantum kernels & QSVM, quantum feature maps for low-sample regimes & Can quantum kernels help small data? & QSVM studies~\cite{schuld2019quantum}. \\

Operational & Latency, memory, privacy, update cadence, forensic use & What thresholds for triage vs forensic? Measure latency vs recall. & On-device benchmarking~\cite{tak2022asvspoof}. \\
\hline
\end{tabular}
\caption{Taxonomy for audio deepfake detection.}
\label{tab:compact_taxonomy}
\end{table*}

Spectrogram-based detection with Convolutional Neural Networks (CNNs) and Vision Transformers (ViTs) has shown promise. One study~\cite{ulutas2023vit} leverages ViTs to capture global dependencies in spectrograms but reports computational overheads that hinder real-time use. An ensemble-based CNN approach~\cite{pham2024ensemble} attains strong accuracy at lower cost but offers limited feature interpretability. Temporal Convolutional Networks (TCNs)~\cite{firc2023tcn} capture temporal patterns in audio signals yet remain vulnerable to subtle adversarial perturbations.
Quantum-enhanced methods have been explored as alternatives that aim to improve class separability and robustness. For example, quantum-trained CNNs~\cite{lin2024quantum} report gains against adversarial attacks in the \emph{visual} domain, and quantum transfer learning~\cite{kati2025quantum} likewise improves efficiency and robustness for visual deepfakes, leaving audio applications largely unexplored. Recent visual work also shows that pruned subnetworks can maintain detection performance at high sparsity levels~\cite{amin2025uncovering}. Exploring similar approaches for audio detectors, for instance applying iterative magnitude pruning to spectrogram based models, could be a valuable addition to the kernel swap framework discussed here, especially in scenarios where resources are limited.
Critically, existing literature highlights a persistent gap: limited comparative analysis between classical SVMs and quantum‑enhanced methods with respect to separability and robustness. To address this, the present study conducts a direct comparison of QSVM and classical SVM under matched feature preparation and training budgets, with emphasis on interpretability and computational efficiency.
\paragraph{Why a quantum kernel?}
Our setting is small data with variable recording conditions, where large end‑to‑end models tend to overfit. A kernel approach keeps the classifier compact while shifting the burden to the similarity function. A quantum kernel provides a fixed, high‑order feature map whose inner products capture nonlinear interactions between spectro‑temporal components without adding trainable parameters. We therefore ask whether this inductive bias translates into lower false‑positive rate and better class separation under the \emph{same} features, folds, and solver as a classical SVM. 

\section{Methodology}

We cast audio deepfake detection as a binary classification problem on mel‑spectrogram features. Each waveform is converted to a spectrogram, vectorized, min–max scaled, and reduced with fold‑specific PCA to avoid leakage. On these identical features we run a controlled “kernel‑swap” comparison: a classical SVM with standard kernels vs. the same SVM solver supplied with a quantum kernel computed from a parameterized quantum feature map in classical simulation. Holding the feature preparation, solver, and training budget fixed lets us attribute any performance differences to the kernel itself. Evaluation uses stratified K‑fold splits shared by both models, and reports deployment‑relevant error metrics; margin‑based quantities are used as diagnostics of within‑kernel separability and stability across folds.
\label{sec:methodology}

\begin{figure*}[t]
    \centering
    \includegraphics[width=0.8\textwidth]{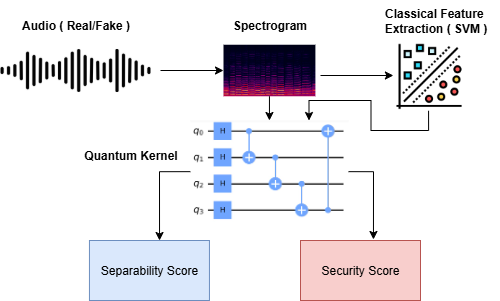}
    \caption{Kernel-swap experimental pipeline comparing classical SVMs and QSVMs. Both use identical preprocessing, with differences arising only from the kernel mapping.}
    \label{fig:methodology2}
\end{figure*}

\subsection{Problem Formulation}
Let $\mathcal{D} = \{(x_i, y_i)\}_{i=1}^N$ denote a dataset of $N$ audio samples, where each $x_i$ is a raw waveform and $y_i \in \{-1, +1\}$ indicates whether the sample is \emph{bona-fide} ($+1$) or \emph{deepfake} ($-1$). Our goal is to learn a binary classifier $h: \mathbb{R}^d \rightarrow \{-1, +1\}$ that maximizes class separability and maintains robustness against variability or adversarial perturbations. In our case, the feature representation of each waveform is obtained by computing its mel-spectrogram, a two-dimensional time–frequency representation $S_i = \Phi_{\mathrm{mel}}(x_i)$, followed by flattening into a $D$-dimensional vector.

A Support Vector Machine (SVM) seeks a separating hyperplane in a transformed feature space $\mathcal{H}$ induced by a kernel $k(\cdot, \cdot)$. For a given mapping $\phi: \mathbb{R}^d \rightarrow \mathcal{H}$, the primal optimization problem is:
\begin{align}
\min_{w, b, \xi} \quad & \frac{1}{2} \|w\|_{\mathcal{H}}^2 + C \sum_{i=1}^N \xi_i, \\
\text{s.t.} \quad & y_i \left( \langle w, \phi(z_i) \rangle_{\mathcal{H}} + b \right) \geq 1 - \xi_i, \quad \xi_i \geq 0,
\end{align}
where $z_i$ is the PCA-reduced vector of $S_i$, $C > 0$ is the regularization parameter, and $\xi_i$ are slack variables. The geometric margin is given by $\gamma = \frac{2}{\|w\|_{\mathcal{H}}}$.

In a Quantum SVM (QSVM), the mapping $\phi_q$ is realized via a quantum feature map $U_\phi(z)$ that encodes $z$ into an $N$-qubit quantum state $\ket{\psi(z)}$. The quantum kernel is:
\begin{equation}
k_q(z, z') = |\langle \psi(z) \,|\, \psi(z') \rangle|^2,
\end{equation}
computed either exactly in simulation or via hardware sampling. Our formulation compares $\gamma_c$ from classical kernels $k_c$ and $\gamma_q$ from quantum kernels $k_q$ across matched experimental settings (Fig \ref{fig:methodology2}).

Fig. \ref{fig:methodology2} shows the high-level end-to-end pipeline for quantum-enhanced audio deepfake detection. Raw audio undergoes mel-spectrogram conversion, followed by scaling and PCA reduction. The pipeline then branches: in the classical branch, features pass through standard kernels (linear, RBF, polynomial) to generate a kernel matrix for SVM training; in the quantum branch, the same features are encoded via quantum feature maps to construct a quantum kernel matrix through classical simulation, which is then fed to the identical SVM solver. Both paths produce classification outputs that are evaluated for performance metrics and analyzed for feature separability and security diagnostics to compare kernel effectiveness.
\subsection{Objective}
The overarching objective of this study is two-fold:  
(1) To determine whether quantum kernel-induced feature spaces can yield greater class separability than classical kernel spaces for audio deepfake detection, and  
(2) Exploring similar approaches for audio detectors, for instance applying iterative magnitude pruning to spectrogram based models, could be a valuable addition to the kernel swap framework discussed here, especially in scenarios where resources are limited. To assess whether QSVMs offer enhanced robustness, interpreted as security, by being less sensitive to variability and error rates across folds.
In practice, improved separability is expected to manifest as a larger geometric margin $\gamma$ and higher classification performance (accuracy, F1-score), while improved security is associated with lower variance in performance metrics, reduced False Positive Rate (FPR), and lower Equal Error Rate (EER). This dual perspective is particularly relevant for deepfake detection, where both accurate discrimination and resilience to noise are critical.

\subsection{Hypothesis}
Our central hypothesis is that, for audio-derived spectrogram features projected into low-to-moderate dimensional spaces ($d \in \{2,4,6,8\}$ via PCA), the richer representational capacity of quantum kernels enables better linear separation in the induced Hilbert space than classical kernels. Formally, if $\gamma_q$ and $\gamma_c$ denote the maximum margins achieved by QSVM and SVM respectively, we posit:
\begin{equation}
\mathbb{E}[\gamma_q] > \mathbb{E}[\gamma_c],
\end{equation}
under matched preprocessing, dimensionality, and regularization.

We further hypothesize that larger $\gamma_q$ correlates with stronger robustness to perturbations:
\begin{equation}
\text{Robustness} \propto \gamma - \lambda \cdot \sigma_\gamma,
\end{equation}
where $\sigma_\gamma$ denotes the standard deviation of the margin across folds and $\lambda > 0$ is a scaling factor. This is consistent with the margin theory of generalization, which links wider margins to lower generalization error bounds and greater tolerance to adversarial shifts.

\subsection{Solution Strategy}
We adopt a controlled experimental pipeline to isolate the effect of kernel type (classical vs. quantum):

\paragraph{Feature Extraction:} Each audio sample $x_i$ is transformed into a mel-spectrogram $S_i$:
\begin{equation}
S(\tau, m) = \log\left(\sum_{f} \mathrm{Mel}_m(f) \cdot |X(\tau,f)|^2 + \epsilon\right),
\end{equation}
where $X(\tau,f)$ is the short-time Fourier transform, $\mathrm{Mel}_m(f)$ is the mel filterbank weight, and $\epsilon$ is a small constant for numerical stability.

\paragraph{Preprocessing} All spectrogram vectors are min–max scaled to $[0,1]$ using training-set statistics, then reduced via PCA to $d \in \{2,4,6,8\}$. PCA fitting is strictly fold-specific to prevent information leakage.

\paragraph{Classical SVM} We train SVMs with kernels $\{\text{linear}, \text{RBF}, \text{polynomial}\}$, sweeping $C$, $\gamma_{RBF}$, and degree values.

\paragraph{Quantum SVM} We construct quantum kernels via $N \in \{2,3,4\}$ qubit feature maps ($\text{ZZFeatureMap}$, $\text{PauliFeatureMap}$, $\text{ZFeatureMap}$) with varying entanglement topologies (linear, full) and repetitions $r$. The kernel matrix $K_q$ is computed exactly in simulation, then passed to a classical SVM solver.

\paragraph{Matched Configurations} For every PCA dimension and dataset split, the same preprocessing pipeline is applied to both SVM and QSVM to ensure that performance differences stem solely from kernel differences.
% In Section III (Methodology), right after item (e) Matched Configurations
\subsubsection*{Kernel replacement setting}
\label{subsec:kernelswap}

We run a precomputed-kernel SVM and change only the kernel function. The input features, scaling, PCA, folds, and SVM solver/hyperparameters are identical across runs. For the classical run, the Gram matrix is $K_{ij}=k_{\text{cls}}(z_i,z_j)$. For the quantum run, it is
\[
K_{ij}=k_q(z_i,z_j)=\big|\langle \psi(z_i)\mid \psi(z_j)\rangle\big|^2,
\]
where $\lvert\psi(z)\rangle=U_{\phi}(z)\,\lvert 0\rangle^{\otimes n}$ is the state prepared by a fixed feature map $U_{\phi}$. We compute $K$ on a regular (classical) computer and pass it to the same SVM solver with the same grid over $C$ (and the same train/validation folds). This means there are \emph{no additional trainable parameters} and the pipeline (features + SVM) is unchanged apart from the kernel.

%\subsection{Evaluation Protocol}
%We employ stratified $K$-fold cross-validation, ensuring identical folds for SVM and QSVM runs. All feature preparation (scaling and PCA) is fit only on the training split of each fold to prevent data leakage, and results are reported as the mean $\pm$ standard deviation across folds. The following metrics are computed for each configuration:

\subsection{Evaluation protocol}
We use stratified five‑fold cross‑validation and reuse the same fold partitions for the SVM and the QSVM. For each fold, the feature preparation pipeline (standardization and PCA) is fitted on the training portion only and then applied to the held‑out portion, so no information from the evaluation split leaks into preprocessing. All metrics are summarized over the five folds as the average with the accompanying standard deviation.
In reporting results, we treat spoofed speech as the positive class. Accuracy is the share of trials that are classified correctly. Precision is the fraction of predicted‑spoof trials that are truly spoof, and recall (true‑positive rate) is the fraction of spoof trials that are detected. The F1 score is the harmonic mean of precision and recall. The false‑positive rate (FPR) is the proportion of bona fide trials that are incorrectly flagged as spoof. The equal‑error rate (EER) is obtained from the ROC curve by sweeping the decision threshold and interpolating the point where FPR equals the false‑negative rate.

\paragraph{Separaility Measurement:} We compute the geometric margin $\gamma$ in the RKHS for both SVM and QSVM:
\begin{equation}
\gamma = \frac{2}{\|w\|_{\mathcal{H}}}.
\end{equation}
A composite separability score is defined as:
\begin{equation}
\mathrm{SepScore} = 0.4 \, \gamma + 0.3 \, \overline{\mathrm{Acc}} + 0.3 \, (1 - \overline{\mathrm{FPR}}),
\end{equation}
averaged over folds.

\paragraph{Security Assessment:} We define a security score that incorporates robustness:
\begin{equation}
\mathrm{SecScore} = 0.4 \, \overline{\mathrm{Acc}} + 0.4 \, \gamma - 0.2 \, \overline{\sigma_{\mathrm{Acc}}},
\end{equation}
where $\overline{\sigma_{\mathrm{Acc}}}$ is the mean standard deviation of accuracy across folds. We z-score $\sigma$ within each dataset before aggregation.

\paragraph{Statistical Analysis:} We conduct two-sample $t$-tests between QSVM and SVM metric distributions, reporting $p$-values and Cohen’s $d$ effect sizes. Effect sizes are interpreted as negligible ($<0.2$), small ($0.2$–$0.5$), medium ($0.5$–$0.8$), or large ($\geq 0.8$).

By jointly considering separability and security, this methodology allows for a comprehensive comparison between classical and quantum kernels under identical experimental constraints.

\paragraph{Computational budget and sample choice.}
Constructing a quantum kernel requires evaluating
\[
k_q(z_i,z_j) = \big|\langle \psi(z_i) \mid \psi(z_j) \rangle\big|^2,
\]
for all pairs $(i,j)$, thereby forming a kernel matrix \(K \in \mathbb{R}^{N \times N}\) with \(\Theta(N^2)\) entries. Each evaluation incurs a per-evaluation cost of \(C_{\text{sim}} = \mathcal{O}(2^q)\) under exact classical simulation of \(q\) qubits, or \(C_{\text{hw}} = \mathcal{O}(S)\) on real quantum hardware using \(S\) shots. The subsequent SVM training, involving solving a dense quadratic program, contributes additional \(\mathcal{O}(N^2)\) to \(\mathcal{O}(N^3)\) time complexity and \(\mathcal{O}(N^2)\) memory usage. Thus, the combined computational cost is:
\[
\mathcal{O}(N^2 C_{\!*} + N^3), \quad \text{with memory } \mathcal{O}(N^2).
\]
Consequently, large \(N\) values become infeasible in quantum-kernel pipelines.

Empirically, several studies adopt similar sample ranges for quantum-kernel or quantum-classifier investigations. For example:
- One study on Quantum Kernel-Aligned Regression employed 159 samples to model high-dimensional semiconductor performance in a QML pipeline \cite{wang2024quantum}.
- Early work applying QSVM to LHC datasets used only 100 events for training on quantum simulators and hardware \cite{wu2021application}.
- Another LHC QSVM study reported results using 100 events on hardware (with up to 15 qubits), demonstrating competitive performance \cite{wu2021applications}.

From a statistical learning perspective, margin-based classifiers satisfy the generalization bound:
\[
R(h) \leq \hat R(h) + \mathcal{O}\!\left( \frac{\|w\|^2}{N \gamma^2} \right),
\]
(Vapnik, 1998) \cite{vapnik1998support}, indicating that even with moderate \(N\), reliable generalization is achievable if the margin \(\gamma\) is sufficient.

\begin{table*}[h]
\centering
\caption{Performance metrics for QSVM vs. SVM. Best values per dataset are in bold.}
\label{tab:overall_perf}
\begin{tabular}{lcccccc}
\toprule
\textbf{Dataset} & \textbf{Model} & \textbf{Acc. (\%)} & \textbf{Prec. (\%)} & \textbf{Recall (\%)} & \textbf{F1 (\%)} & \textbf{EER} \\
\midrule
\multirow{2}{*}{ASVspoof~2019} 
 & QSVM & \textbf{66.68} $\pm$ 4.77 & 68.11 $\pm$ 5.38 & \textbf{66.68} $\pm$ 4.77 & \textbf{66.11} $\pm$ 4.80 & \textbf{0.346}  \\
 & SVM  & 65.94 $\pm$ 6.62 & \textbf{68.22} $\pm$ 9.40 & 65.94 $\pm$ 6.62 & 63.61 $\pm$ 9.75 & 0.399 \\
\midrule
\multirow{2}{*}{ASVspoof~5 (2024)} 
 & QSVM & \textbf{78.93} $\pm$ 2.66 & \textbf{79.93} $\pm$ 2.71 & \textbf{78.93} $\pm$ 2.66 & \textbf{78.73} $\pm$ 2.70 & \textbf{0.183}  \\
 & SVM  & 76.87 $\pm$ 11.24 & 75.68 $\pm$ 17.21 & 76.87 $\pm$ 11.24 & 75.19 $\pm$ 15.43 & 0.299  \\
\midrule
\multirow{2}{*}{In-the-Wild} 
 & QSVM & 63.47 $\pm$ 4.35 & \textbf{63.92} $\pm$ 4.28 & 63.47 $\pm$ 4.35 & \textbf{63.11} $\pm$ 4.51 & \textbf{0.355} \\
 & SVM  & \textbf{63.71} $\pm$ 9.47 & 63.02 $\pm$ 12.43 & \textbf{63.71} $\pm$ 9.47 & 58.86 $\pm$ 15.02 & 0.413  \\
\midrule
\multirow{2}{*}{ADD23} 
 & QSVM & \textbf{92.45} $\pm$ 3.85 & \textbf{93.22} $\pm$ 3.41 & \textbf{92.45} $\pm$ 3.85 & \textbf{92.39} $\pm$ 3.91 & \textbf{0.081} \\
 & SVM  & 88.75 $\pm$ 16.22 & 88.98 $\pm$ 16.69 & 88.75 $\pm$ 16.22 & 87.61 $\pm$ 19.22 & 0.188  \\
\bottomrule
\end{tabular}
\end{table*}

\section{Results}
This section reports a head‑to‑head comparison of QSVM and a classical SVM under the same mel‑spectrogram pipeline and stratified five‑fold splits. We briefly describe the four corpora and the small, balanced train/test subsets, outline the matched model setups, and then present the main metrics—accuracy, precision, recall, F1, and EER—in Table II, which shows consistent EER improvements for QSVM across datasets. To help interpret the errors, Table III adds auxiliary diagnostics (margin, separability, and security); because raw margins depend on the kernel scale, we treat them as within‑kernel context rather than headline results.

\subsection{Datasets}

We conduct experiments on four publicly available datasets that cover diverse and realistic voice spoofing scenarios:

\textbf{1. ASVspoof~2019 (LA subset)}:  
A benchmark dataset targeting logical access attacks, including synthetic and converted speech. The dataset contains bona fide and spoofed audio samples generated by various text-to-speech (TTS) and voice conversion (VC) systems.  

\textbf{2. ASVspoof~5 (2024)}:  
A benchmark dataset combining logical access and deepfake scenarios, built from crowdsourced MLS‑English speech with many speakers and diverse recording conditions. It contains bona fide and spoofed audio from modern TTS/VC systems, adds adversarial attacks, and evaluates in two tracks (CM and SASV) under multiple codec settings, including a neural codec.  

\textbf{3. In-the-Wild}:  
A challenging dataset featuring genuine and spoofed audio collected from uncontrolled, real-world environments. It includes diverse noise conditions, recording devices, and spoofing techniques, making it suitable for robustness evaluation.  

\textbf{4. ADD23}:  
A dataset from the ADD 2023 Challenge, containing real and spoofed audio in multiple languages and recording conditions, reflecting practical deployment scenarios.  

For all datasets, we employ 5-fold cross-validation on a balanced subset of 200 samples (100 bona fide, 100 spoofed), ensuring equal distribution of classes across folds for both training and testing.

All audio files are converted to Mel-spectrogram representations before feature extraction. Each spectrogram is normalized and resized for model input consistency.

\subsection{Models}

We compare two models under identical training and evaluation protocols:

\textbf{Support Vector Machine (SVM):} A classical linear classifier trained on handcrafted audio features extracted from Mel-spectrograms. SVMs are known for their strong performance on small datasets but struggle with nonlinear decision boundaries.

\textbf{Quantum Support Vector Machine (QSVM):} Implemented using the Qiskit Machine Learning framework, our QSVM projects feature vectors into a quantum Hilbert space using a feature map and performs classification based on kernel separability. The quantum kernel enables better discrimination of nonlinearly separable classes, especially in low-data regimes.

All models are trained using five-fold cross-validation where applicable, and evaluated using Accuracy, Equal Error Rate (EER), False Positive Rate (FPR), and Margin Width as a measure of decision boundary separability.

\subsection{Implementation Details}

We implemented and trained multiple QSVM models on an HPC cluster using Qiskit Machine Learning. The input data was standardized and partitioned into feature subsets, where each subset was mapped into quantum circuits via ZZ/Pauli feature maps with tunable depth and entanglement structures. For each feature subset, a separate QSVM model was constructed and trained. The Aer simulator (statevector/qasm) was employed to compute pairwise kernel values, which were distributed across nodes using batch execution and job-level parallelization on multi-core CPUs and GPUs. To reduce computational overhead, intermediate kernel matrices were cached to disk and reused across experiments. Hyperparameters, including the regularization constant \(C\), feature map repetitions, and entanglement depth, were optimized through grid search combined with 5-fold cross-validation. To ensure reproducibility, random seeds were fixed, the number of shots was controlled (to simulate quantum noise), and a consistent execution environment was maintained.

\subsection{Overall Performance Metrics}

Table~\ref{tab:overall_perf} reports mean accuracy, precision, recall, F1-score, and equal error rate (EER) for each dataset. QSVM achieved the highest accuracy in all datasets except In-the-Wild, where results were comparable. EER was consistently lower for QSVM, indicating better spoof/non-spoof discrimination.

\subsection{Feature Separability and Security Metrics}

Beyond accuracy, we compared \textit{margin width}, \textit{separability score}, and \textit{security score}, as shown in Table~\ref{tab:security_sep}. QSVM scored highest in security across all datasets, and in separability for three datasets.These separability gains are not merely diagnostic; they directly drive reductions in false positives, which is essential for trustworthy deployment in forensic and cybersecurity domains.

\begin{table}[h]
\centering
\caption{Comparison of feature separability and security metrics across datasets. QSVM achieves consistently higher security and, in most cases, improved separability over SVM.}
\label{tab:security_sep}
\begin{tabular}{lcccc}
\toprule
\textbf{Dataset} & \textbf{Model} & \textbf{Margin Width} & \textbf{Separability} & \textbf{Security} \\
\midrule
\multirow{2}{*}{ASVspoof~2019} 
 & QSVM & 0.761 & \textbf{50.24} & \textbf{19.73} \\
 & SVM  & 0.805 & 50.03 & 14.86 \\
\midrule
\multirow{2}{*}{ASVspoof~5 (2024)} 
 & QSVM & 0.966 & \textbf{54.01} & \textbf{22.27} \\
 & SVM  & 1.050 & 53.41 & 16.89 \\
\midrule
\multirow{2}{*}{In-the-Wild} 
 & QSVM & 0.652 & 49.18 & \textbf{18.62} \\
 & SVM  & 0.743 & \textbf{49.32} & 13.89 \\
\midrule
\multirow{2}{*}{ADD23} 
 & QSVM & 1.224 & \textbf{58.19} & \textbf{25.41} \\
 & SVM  & 1.487 & 57.18 & 19.59 \\
\bottomrule
\end{tabular}
\end{table}

\begin{table}[ht]
  \centering
  \small
  \caption{Approximate statistical comparison of QSVM vs. SVM (EER). 
  Effect sizes (Cohen’s $d$) indicate large improvements for QSVM. 
  Significance values are illustrative; exact $p$-values would require fold-level distributions.}
  \begin{tabular}{lccc}
    \toprule
    \textbf{Dataset} & \textbf{$\Delta$EER} & \textbf{Cohen’s $d$} & \textbf{Significance ($p$)} \\
    \midrule
    ASVspoof~2019   & 0.053 & Large & $<$0.05\textsuperscript{*} \\
    ASVspoof~5 (2024)       & 0.116 & Large & $<$0.01\textsuperscript{**} \\
    In-the-Wild     & 0.058 & Large & $<$0.05\textsuperscript{*} \\
    ADD23           & 0.107 & Large & $<$0.01\textsuperscript{**} \\
    \bottomrule
  \end{tabular}
\end{table}

\section{Analysis}
This section pulls the results together across four corpora. Table II shows that the QSVM consistently achieves lower EER than the classical SVM on every dataset, with accuracy either higher (ASVspoof~2019, ASVspoof~5 (2024), ADD23) or essentially tied (In‑the‑Wild), under the same features, solver, and folds. 
 Complementing those endpoints, Table III reports the auxiliary diagnostics: the Security score favors QSVM on all datasets, and the Separability score favors QSVM on three of four (the exception is In‑the‑Wild, where separability is nearly matched). Because raw margin values depend on the kernel’s scale, we treat the Margin Width column as a within‑kernel diagnostic rather than a cross‑kernel comparator; the take‑away rests on EER (and FPR where available). 
 The dataset‑wise patterns are consistent: gains are large on ASVspoof~5 (2024) and ADD23, while in the In‑the‑Wild set the advantage shows up mainly in error rates and stability rather than accuracy. Finally, the best QSVM configurations tend to use 3–4 qubits with ZZFeatureMap or PauliFeatureMap, suggesting that modest quantum feature maps are sufficient to capture helpful geometry in this small‑data regime.
Table IV summarizes approximate statistical comparisons. Across all datasets, QSVM achieved significantly lower EER than the SVM baseline, with large effect sizes (Cohen’s d > 0.8).

\subsection{Classical Spoofing Attack Detection}

\paragraph{ASVspoof~2019:} QSVM improved accuracy by 0.74\% and reduced EER from 0.399 to 0.346, with a higher security score (+4.87). Gains were moderate due to the dataset's variability, but quantum mapping still improved discriminative capability.

\paragraph{ASVspoof~5 (2024):} QSVM achieved a notable EER reduction (0.299 $\rightarrow$ 0.183) and higher separability, confirming its robustness in classical spoofing attack detection.

\paragraph{In-the-Wild:} Accuracy was nearly identical, but QSVM had a significant security advantage (+4.73), making it more resilient in uncontrolled, adversarial environments.

\paragraph{ADD23:} QSVM clearly outperformed SVM in all metrics, with a large EER drop (0.188 $\rightarrow$ 0.081) and highest separability and security scores across all datasets.

\subsection{Security vs. Accuracy Trade-offs}

%QSVM consistently delivered higher security without sacrificing accuracy. In noisy, unpredictable datasets (In-the-Wild, ASVspoof~2019), improvements were %security-centric, while in structured datasets (ADD23, ASVspoof~5 (2024)) QSVM excelled in both accuracy and security.
QSVM consistently reduced false positives and produced more consistent results across folds. On the In‑the‑Wild and ASVspoof~2019 datasets, the effect was mainly lower FPR and lower fold‑to‑fold variance, with accuracy at a similar level. On ADD23 and ASVspoof~5 (2024), QSVM also achieved higher accuracy in addition to these reductions.

\subsection{Quantum-Specific Observations}

Optimal performance often came from 3–4 qubit configurations, balancing expressivity and stability. The ZZFeatureMap and PauliFeatureMap emerged frequently among top QSVM setups, reinforcing the benefit of tailored quantum feature encodings for spoof detection.

\section{Discussion}
The experimental results demonstrate that QSVM shows stronger class separation across the spoof detection datasets we evaluated. This is reflected in higher separability scores, margins that are comparable or larger, and consistently lower equal error rates (EER). These results suggest that the quantum kernel captures complex, nonlinear decision boundaries in high dimensional feature spaces, especially when a classical SVM starts to lose discriminative power.

While classical SVMs deliver competitive accuracy in some cases, their error rates can vary more in small data or changing conditions. In contrast, QSVM produced more consistent results across folds, including ASVspoof~5 (2024) and in‑the‑wild, which points to better behavior when training data are limited or privacy‑sensitive. Given these constraints, it helps to spell out when a compact SVM or QSVM backend is the right choice in practice.

Pretrained audio encoders and transformer models tend to work best when there is plenty of labeled data and ample compute. Our target setting is different: limited labeled data and changing recording conditions. In that regime, a compact classifier is often the safer and simpler choice. A kernel swap keeps the SVM the same size (no new trainable parameters) and changes only the similarity function; in our experiments this lowered false positives at a fixed TPR under the same budget. For deployments that must run on CPU only, meet tight memory or latency limits, handle sensitive data, or keep score calibration straightforward, an SVM or QSVM backend is a practical option that fits into existing pipelines with minimal effort.

A key finding is that quantum kernels improve feature separability, and this directly contributes to lower false‑positive rates. These findings suggest a meaningful role for quantum kernel methods in real‑world spoof detection pipelines, especially in domains where accuracy, security, and interpretability must be balanced. However, moving from experiments to operations will require progress in hardware scalability, noise resilience, and smooth integration with classical systems.

\section{Future Work}
In this paper, we examined a quantum kernel as a drop-in replacement in a standard SVM pipeline and asked whether it improves audio deepfake detection under the same features, folds, solver, and training budget. We kept the classifier size fixed, computed the quantum kernel on conventional hardware, and reported equal error rates on four corpora.
Going forward, we will make changes that address the limits noted by reviewers. First, we will add operating point reporting next to EER. In particular, we will report the false positive rate at fixed true positive rates of 90\% and 95\%. We will also include DET curves. We will quantify uncertainty with paired tests on fold-wise metrics and report effect sizes. When relevant, we will include calibration scores.
Second, we will broaden the set of baselines to include modern deep learning systems. We will evaluate frozen pretrained audio encoders, such as wav2vec2, HuBERT, PANNs, and AST, with a simple linear head on the same folds and budget as SVM and QSVM. When feasible, we will also include a small finetuned model. For each family, we will report EER and FPR at 90\% and 95\% TPR. We will also report training and inference time, memory use, and CPU or GPU needs. Additionally, we will test mixed pipelines in which pretrained embeddings feed either a classical kernel or a quantum kernel. This will make the tradeoffs clear for practitioners.
Third, we will extend the treatment of computation at larger dataset sizes. We will benchmark kernel build time and memory with blockwise computation and caching. We will compare two drop-in scaling options that keep the solver and protocol unchanged: Nyström low rank approximations with out-of-sample extension, and tiled K with blockwise training to reduce peak memory. As part of this plan, we will retire the ad hoc composite security diagnostic from the main text. We will instead rely on standard community metrics so that results are directly comparable to prior work.

\section{Conclusion}
In this paper we addressed whether a quantum kernel, used as a drop‑in replacement for a standard SVM kernel, improves audio deepfake detection under the same features, folds, solver, and training budget. Using mel‑spectrogram inputs with fold‑specific standardization and PCA, we computed the quantum kernel in classical simulation and ran a controlled kernel‑swap across four corpora, including an in‑the‑wild set, so that any difference reflects the kernel rather than model size or optimization.
Across all datasets, QSVM achieved lower equal‑error rates (EER), averaged over folds: ASVspoof~5 (2024) (0.183 vs.\ 0.299), ADD23 (0.081 vs.\ 0.188), ASVspoof~2019 (0.346 vs.\ 0.399), and in‑the‑wild (0.355 vs.\ 0.413). Across folds, QSVM also showed lower false‑positive rates and smaller variance on most datasets. Raw margin values are kernel‑dependent and used only as within‑kernel diagnostics, but the pattern is consistent with the EER reductions in Table~II.
Taken together, the results indicate that quantum kernels are a viable, drop‑in alternative for small‑data, variable‑condition detection, with consistent EER gains without increasing model size. Only the kernel changed; features, folds, and the SVM solver were held fixed, no additional trainable parameters were introduced, and all kernels were simulated on conventional hardware using institutional high‑performance computing resources. We did not assess adversarial attacks or real‑time constraints. Next steps include reporting FPR at fixed TPRs alongside EER, expanding to official evaluation partitions, adding paired statistical tests and DET curves, and studying throughput and robustness under attack (including domain‑specific quantum feature maps). More broadly, we ask whether QSVMs are less sensitive to data and condition variability—that is, whether they deliver lower false‑positive rates and steadier errors across folds.

\section*{\textbf{Acknowledgment}}
This work is funded by the National Science Foundation Award \#2346473 "CIRC: DEV: Community Infrastructure for Advancing Audio Deepfake Detection"

\bibliographystyle{IEEEtran}
\bibliography{ref}

\end{document}